\documentclass[11pt]{article}

\usepackage{amsmath,amssymb,amsthm,setspace,graphicx}
\usepackage[text={6.5in,8.4in}, top=1.3in]{geometry}
\usepackage{hyperref}

\newtheorem{theorem}{Theorem}[section]

\newtheorem{conjecture}[theorem]{Conjecture}

\newtheorem{definition}[theorem]{Definition}

\renewcommand{\L}{{\cal L}}
\newcommand{\Q}{{\cal Q}}
\newcommand{\T}{{\cal T}}

\begin{document}

\title{On the quartet distance given partial information}

\author{
Sagi Snir
\thanks{Department of Evolutionary Biology, University of Haifa, Haifa 3498838, Israel (ssagi@research.haifa.ac.il).}
\and
Osnat Weissberg
\thanks{Department of Evolutionary Biology, University of Haifa, Haifa 3498838, Israel (osnat.weissberg@gmail.com).}
\and
Raphael Yuster
\thanks{Department of Mathematics, University of Haifa, Haifa 3498838, Israel
(raphael.yuster@gmail.com). The author's research is supported in part by ISF grant 1082/16.}
}

\date{}

\maketitle

\setcounter{page}{1}

\begin{abstract}
	Let $T$ be an arbitrary phylogenetic tree with $n$ leaves. It is well-known that the average quartet distance between two assignments of taxa to the leaves of $T$ is $\frac 23 \binom{n}{4}$.
	However, a longstanding conjecture of Bandelt and Dress asserts that $(\frac 23 +o(1))\binom{n}{4}$ is also the {\em maximum} quartet distance between two assignments.
	While Alon, Naves, and Sudakov have shown this indeed holds for caterpillar trees,
	the general case of the conjecture is still unresolved. A natural
	extension is when partial information is given: the two assignments are known to coincide on a given subset of taxa. The partial information setting is biologically relevant as the location of some taxa (species) in the phylogenetic tree may be known, and for other taxa it might not be known.
	What can we then say about the average and maximum quartet distance in this more general setting? Surprisingly, even determining the {\em average} quartet distance becomes a nontrivial task in the partial information setting and determining the maximum quartet distance is even more challenging,
	as these turn out to be dependent of the structure of $T$. In this paper we prove nontrivial asymptotic bounds that are sometimes tight for the average quartet distance in the partial information setting. We also show that the Bandelt and Dress conjecture does not generally hold under the partial information setting. Specifically, we prove
	that there are cases where the average and maximum quartet distance substantially differ.
	
	\vspace*{3mm}
\end{abstract}

\section{Introduction}

Phylogenetic tree reconstruction is a central problem in computational biology.
An important concept in this area are undirected phylogenetic trees, where the taxa (species) are mapped bijectively to the leaves of the tree and the tree-structure represents the evolutionary
relationship among the taxa. More formally, a phylogenetic tree $T$ of order $n$
is an undirected tree with $n$ leaves where each internal vertex is incident with three edges.
For a taxa set $[n]=\{1,\ldots,n\}$, a bijection between $[n]$ and the leaves of $T$ forms a
{\em labeled} phylogenetic tree. Hereafter, when we refer to trees or labeled trees we always
mean phylogenetic trees or labeled phylogenetic trees, respectively.

\begin{figure}
	\includegraphics[scale=0.4,trim=-190 250 0 130, clip]{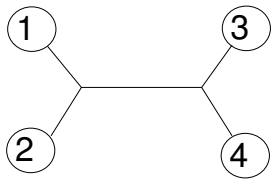}
	\caption{The quartet tree $[12|34]$.}
	\label{f:quartet}
\end{figure}

The smallest labeled tree is a {\em quartet}, a tree with four leaves.
A quartet with leaves labeled $\{a,b,c,d\}$ is denoted by $[ab|cd]$ when there is an edge in the tree separating the pair
$\{a,b\}$ from the pair $\{c,d\}$, as shown in Figure \ref{f:quartet}.
Observe that there are precisely three possible labeled quartets on a given set of four taxa $\{a,b,c,d\}$:
either $[ab|cd]$ or $[ac|bd]$ or $[ad|bc]$.
A labeled tree of order $n$ therefore contains the information of exactly $\binom{n}{4}$ quartets.

Significant research effort has been devoted to phylogenetic tree reconstruction from quartets, a field denoted as {\em quartet-based reconstruction } (see
e.g. \cite{BS-2001,grunewald-2012,GHS-2008,Jiang-SICOMP-2000,Puzzle}).
Here, several (or even all $\binom{n}{4}$) quartets are constructed,
normally using some {\em quartet oracle}~\cite{Csuros-JCB-2002,Daskalakis-SIDMA-2011,Gronau-RSA-2012}. These quartets are combined together into the big tree on the full taxa set.
The  problem of deciding whether there exists a tree compatible\footnote{The notion of compatibility, which is central in phylogeny, is formally defined in Section 2.} to all the quartets in a given set, is NP-complete \cite{steel-1992}. Moreover, the ideal case in which all
quartets are compatible with a single tree, is very rare.
This raises the problem of finding a tree maximizing the number of
compatible quartets - {\em maximum quartet compatibility} (MQC) \cite{SSbook03}.
As MQC is NP-hard, several approximation algorithms have been designed for it
\cite{Bryant-TCS-2006,Jiang-FOCS-98,Snir-SIDMA-2011,Snir-SICOMP-2012},
but the best approximation to the general problem is still obtained by a naive ``random tree'' with expected approximation ratio of $\frac{1}{3}$. 

Related to the problem of compatibility is the concept of \emph{quartet	distance} \cite{EMM-1985,SP-1993}.
This notion quantifies the similarity of two different labeled trees of order $n$, or even two identical trees with 
different labellings, by counting how many quartets are compatible with both of them.
More formally, if $T_1$ and $T_2$ are
two labeled trees of order $n$, let $qd(T_1,T_2)$ denote the difference
between $\binom{n}{4}$ and the number of quartets compatible with both
$T_1$ and $T_2$. Several questions immediately arise. Here are some notable ones:
How large can $qd(T_1,T_2)$ possibly be?
How large could it be if $T_1$ and $T_2$ have the same tree structure?
In the latter case, how large could it be if we know that some subset of taxa are pinned to their leaves in both trees? What can be said about (the random variable) $qd(T_1,T_2)$ if $T_1$ and $T_2$ are the same tree, and a subset of taxa are pinned to their leaves in both trees, while the unpinned taxa are placed randomly in the remaining leaves in both trees?

Somewhat surprisingly, already for the first question, the answer is significantly smaller than $\binom{n}{4}$, namely, one can never obtain total
incompatibility. Bandelt and Dress \cite{BD-1986} showed that the maximum is always strictly
smaller than $\frac{14}{15}\binom{n}{4}$ for $n \ge 6$. They also conjectured
that the ratio between the maximum quartet distance and $\binom{n}{4}$
converges to $\frac 23$ as $n$ tends to infinity.
\begin{conjecture}[Bandelt and Dress]
\label{conj:bd}
The maximum quartet distance between two labeled trees of order $n$ is
$\left(\frac 23 + o(1)\right) \binom{n}{4}$.
\end{conjecture}
As the lower bound of $\frac 23 \binom{n}{4}$ can
be easily obtained by the random labeling argument (there are also highly explicit constructions
achieving this lower bound \cite{CEK-2019}),
Conjecture \ref{conj:bd} implies, perhaps surprisingly, that the average distance between two random
trees is asymptotically the same as the maximum distance.
Alon, Snir, and Yuster \cite{ASY-2014} improved the upper bound 
on the maximum quartet distance proving it is asymptotically smaller
than $\frac{9}{10} \binom{n}{4}$. Alon, Naves, and Sudakov \cite{ANS-2016} further improved the
upper bound to at most $\left(0.69 + o(1)\right) \binom{n}{4}$.
In fact, for the particular important case where both $T_1$ and $T_2$ are the
{\em caterpillar tree} they established the validity of Conjecture \ref{conj:bd}.
We stress that even if $T_1$ and $T_2$ are assumed to be different labellings of the {\em same} tree, nothing
better than the lower bound ratio $\frac{2}{3}+o(1)$ and the upper bound $0.69+o(1)$ is known.

We note that the random labeling which achieves the $\frac{2}{3}$ ratio is the natural approach one takes
when no information on the tree or its labeling is given.
In fact, even if the tree structure is known, but no information on the labeling is given, the natural approach is still random labeling. However, a more likely scenario is that {\em some information} is known.
Our contribution here is for this more general case. We consider here the case where the tree
structure is known and some subset of taxa are known to be at certain leaves.
It is reasonable to expect that the more information given, the random labeling approach of the remaining taxa
will achieve a better ratio, approaching $0$ when almost full information is given.
However, quantifying this ratio, while straightforward in the ``no information'' setting, becomes difficult
in the partial information setting. Likewise, it is reasonable to expect that the more information given,
the maximum possible distance will achieve a better ratio. Quantifying this ratio is not only difficult, but it is also not clear what should be the ``right conjecture''. In fact, in some cases, it is provably not true that
the average distance is asymptotically the same as the maximum distance, as conjectured by Bandelt and Dress in the no information setting.

In the next section we formalize the aforementioned problems and results whereas in the later sections we prove our main claims.

\section{Problems definitions and statements of main results}

For a tree $T$, the set of leaves of $T$ is denoted by $\L(T)$.
For $A \subseteq \L(T)$, the topological subtree
$T|_A$ is the tree obtained by removing the leaves $\L(T)\setminus A$,
the paths leading exclusively to them,  and contracting vertices with degree two.

Recall that an isomorphism $\simeq$ between two graphs where some of the vertices of the graphs are labeled and some are not is a bijective function between the vertices which preserves adjacencies and non-adjacencies and a labeled vertex in the first tree is mapped to a labeled vertex in the other tree having the same label, while a
non-labeled vertex in the first tree is mapped to a non-labeled vertex in the other tree.
For two labeled trees $T$ and $T'$, we say that $T$ {\em satisfies}
$T'$ if $\L(T') \subseteq \L(T)$ and $T|_{\L(T')} \simeq T'$, namely
the topological subtree of $T$ induced by $\L(T')$ is isomorphic to $T'$.

If $T'$ is a quartet, we say that $T'$ is {\em compatible} with $T$ if $T$ satisfies $T'$.
Observe that if $T$ is a labeled tree of order $n$, then it contains precisely $\binom{n}{4}$
compatible quartets, each one corresponding to the labeled topological subtree obtained by considering four taxa.
Let, therefore $\Q(T)$ denote the quartet set consisting of all compatible quartets of the labeled tree $T$.
Now, given two labeled trees $T_1$ and $T_2$ of order $n$, (that is, both are labeled with the taxa set $[n]$)
let $qd(T_1,T_2)=\binom{n}{4} - |\Q(T_1) \cap \Q(T_2)|$. We call $qd(T_1,T_2)$ the {\em quartet distance} between
$T_1$ and $T_2$. We mention here that Colonius and Schulze observed that
if $qd(T_1,T_2)=0$ then $T_1$ and $T_2$ are isomorphic (see \cite{SSbook03}). 
Of course, it may be that $T_1$ and $T_2$ have the same tree-structure (i.e. they are isomorphic as unlabeled trees) but $qd(T_1,T_2) > 0$, for example $qd([ab|cd],[ac|bd])=\binom{4}{4}-0=1$.

Let $\T_n$ denote the set of all labeled trees of order $n$. Consider the equivalence relation on
$\T_n$ where two elements are equivalent if they have the same tree structure. Hence, we can
represent an equivalence class by $T$, where $T$ is an unlabeled tree of order $n$.
For example, if $n=4$ then there is only one equivalence class (an unlabeled quartet) consisting of
the three possible quartets on $\{1,2,3,4\}$. Let $X_n$ be the random variable corresponding to
$qd(T_1,T_2)$ where $T_1$ and $T_2$ are uniformly drawn (say, with replacement) from $\T_n$.
By linearity of expectation we have that $\mathbb{E}[X_n]=\frac{2}{3}\binom{n}{4}$.
Similarly, for an unlabeled tree $T$ of order $n$, let $X_T$ be the random variable corresponding to
$qd(T_1,T_2)$ where $T_1$ and $T_2$ are uniformly drawn random labellings of $T$ (one can view this as
sampling from $\T_n$ conditioned on a given tree structure).
Again, linearity of expectation gives $\mathbb{E}[X_T]=\frac{2}{3}\binom{n}{4}$.
Let $M_n$ be the maximum possible value that $X_n$ can attain, or stated otherwise, just the maximum possible quartet distance between two elements of $\T_n$. Similarly, let $M_T$ be the maximum possible value that $X_T$ can attain, or stated otherwise, the maximum possible quartet distance between two labellings of $T$.
Notice that $M_n \ge M_T$ for every tree $T$ of order $n$.

Using the notations $X_n,X_T,M_n,M_T$ we can now reformulate the aforementioned results and conjecture from the introduction. Bandelt and Dress conjectured that
$M_n = (1+o(1))\frac{2}{3}\binom{n}{4}=(1+o(1))\mathbb{E}[X_n]$ which, if true, implies that
$M_T = (1+o(1))\frac{2}{3}\binom{n}{4}=(1+o(1))\mathbb{E}[X_T]$ for every tree $T$ of order $n$.
The result of \cite{ANS-2016} proves that $M_n \le (0.69+o(1))\binom{n}{4}$.
The {\em caterpillar} ${\rm CAT}_n$ of order $n \ge 4$ is the tree having only two internal vertices $x,y$ adjacent to two leaves each. We call the four edges connecting $x$ to its two adjacent leaves
and $y$ to its two adjacent leaves {\em endpoint edges}.
So, a quartet is a caterpillar of order $4$ where every edge except the central edge is an
endpoint edge and every tree of order $5$ is isomorphic to ${\rm CAT}_5$, but already for $n=6$ not every tree is a caterpillar. The result of \cite{ANS-2016} also proves that
$M_{{\rm CAT}_n} = (1+o(1))\frac{2}{3}\binom{n}{4}$ so at least for caterpillars, the Bandelt and Dress conjecture holds.

We are now ready to state our generalization for $X_T$ and $M_T$ to the partial information setting.
Let $T$ be an unlabeled tree of order $n$ and
let $S \subseteq \L(T)$ be a subset of leaves. An {\em $S$-labeling} (also called a partial labeling if $S$ is clear from the context) is a labeling of $S$ with the taxa set $[k]$ where $k=|S|$.
We think of a partial labeling as if some information is given to us: we are told which leaf is assigned with a taxa from $[k]$ but we have no information about the assignment of the remaining $n-k$ taxa to the remaining $n-k$ leaves. As we are concerned with asymptotic behavior, we set $k=\alpha n$ where
$\alpha \in [0,1]$ is a fixed rational and always assume that $n$ is such that $\alpha n$ is an integer.
We let $T(S)$ denote the tree $T$ with its $S$-labeling.
A {\em completion} of $T(S)$ is a full labeling of $T(S)$ with the remaining $n-k$ taxa assigned to the remaining set of leaves $\L(T) \setminus S$.
Let $X_{T(S)}$ be the random variable corresponding to $qd(T_1,T_2)$ where $T_1$ and $T_2$ are
independently and randomly chosen completions of $T(S)$.
Notice that it is more convenient and equivalent to view $T_1$ as some fixed completion
and only consider $T_2$ as  randomly chosen from all $(n-k)!$ possible completions.
Let $M_{T(S)}$ be the maximum possible quartet distance between two completions of $T(S)$
(again, it may be more convenient and equivalent to fix one completion and consider the maximum quartet distance between the fixed completion and the other completions).
Notice that $X_T=X_{T(\emptyset)}$, $M_T=M_{T(\emptyset)}$ and $X_{T(\L(T))}=M_{T(\L(T))}=0$.

The definitions of $X_{T(S)}$ and $M_{T(S)}$ as generalizations of the well-studied ``no-information''
variants $X_T$ and $M_T$ give rise to several intriguing questions from which naturally stand out the following, where we assume that $\alpha \in [0,1]$ is fixed, $T$ is an unlabeled tree of order $n$,
and $S \subset \L(T)$ has $|S|=\alpha n$.
\begin{definition}
\begin{align*}
	minexp(T,\alpha) & = \min_{S} {\mathbb E}[X_{T(S)}]\\
	minexp(n,\alpha) & = \min_{T} minexp(T,\alpha)\\
	maxexp(T,\alpha) & = \max_{S} {\mathbb E}[X_{T(S)}]\\
	maxexp(n,\alpha) & = \max_{T} maxexp(T,\alpha)\\
	maxqd(T,\alpha)~\, & = \max_{S} M_{T(S)}\\
	maxqd(n,\alpha)~\, & = \max_{T}  maxqd(T,\alpha)
\end{align*}
\end{definition}
Hence, assuming an $\alpha$ fraction of the taxa are bound to fixed locations,
$minexp(T,\alpha)$ (resp. $maxexp(T,\alpha)$) asks for the smallest (resp. largest) expectation of the quartet distance of two labellings of $T$,
and $minexp(n,\alpha)$ (resp. $maxexp(n,\alpha)$) asks for the smallest (resp. largest) expectation of the quartet distance of two labellings of any tree of order $n$.
Likewise, $maxqd(T,\alpha)$ asks for the largest quartet distance of two labellings of $T$ and $maxqd(n,\alpha)$ asks for the largest quartet distance of two labellings of any tree of order $n$.

When $\alpha = 1$, all of these parameters are zero. On the other hand, when $\alpha=0$ all the first
four parameters are (trivially) $\frac{2}{3}\binom{n}{4}$ and the fifth parameter amounts to $M_T$. But when
$0 < \alpha < 1$ all the first four parameters (moreover the fifth and sixth) become nontrivial to determine.

As there are exponentially many non-isomorphic
trees of order $n$, it seems hopeless to obtain a ``closed'' formula for
$minexp(T,\alpha)$, $maxexp(T,\alpha)$,
and $maxqd(T,\alpha)$ for each possible $T$. Hence, the main combinatorial challenge is to either determine or at least find nontrivial bounds for $minexp(n,\alpha)$, $maxexp(n,\alpha)$ and $maxqd(n,\alpha)$.

We next state our main results which concern $minexp(n,\alpha)$, $maxexp(n,\alpha)$.
We mention that the constants in the forthcoming theorems are (quite obviously) polynomials
of degree $4$ in $\alpha$. For clarity of exposition we will therefore state our results also for the
median case $\alpha=\frac{1}{2}$.
Our first result concerns $minexp(n,\alpha)$ which is asymptotically determined.
\begin{theorem}\label{t:minexp-n}
$$
minexp(n,\alpha) = \left(\frac{2}{3}(1-\alpha)^4+\frac{8}{3}\alpha(1-\alpha)^3\right)\binom{n}{4}(1+o(1))
\;.
$$
In particular,
$$
minexp(n,{\textstyle \frac{1}{2}})=\frac{5}{24}\binom{n}{4}(1+o(1))\;.
$$
\end{theorem}

The method used in the proof of Theorem \ref{t:minexp-n}, which suffices to asymptotically determine
$minexp(n,\alpha)$ can also be used to supply a relatively simple upper bound of $\frac{5}{8}\binom{n}{4}(1+o(1))$ for $maxexp(n,\frac{1}{2})$
and, more generally, $\left(\frac{2}{3}-\frac{2\alpha^4}{3}\right)\binom{n}{4}+ o(n^4)$
for $maxexp(n,\alpha)$ (see Section 4). However as we next prove, this upper bound is not tight and one can always do significantly better. Furthermore, the lower bound we obtain for $maxexp(n,\alpha)$, is quite close to it. The proof of our upper bound requires using linear and non-linear optimization.
\begin{theorem}\label{t:maxexp-n}
$$
\left(\frac{2}{3}-\frac{653\alpha^4+414\alpha^3+33\alpha^2}{1650}\right)\binom{n}{4}+o(n^4)
$$
$$
\ge maxexp(n,\alpha) \ge
$$
$$
\left(\frac{2}{3}-\frac{10\alpha^4+32\alpha^3+12\alpha^2}{81}\right)\binom{n}{4}+o(n^4)\;.
$$
In particular,
$$
0.606\binom{n}{4}(1+o(1)) \ge maxexp(n,{\textstyle \frac{1}{2}}) \ge 0.572\binom{n}{4}(1+o(1))\;.
$$
\end{theorem}

As for $maxqd(n,\alpha)$, the situation becomes even more involved bearing in mind that $maxqd(n,0)$ is a longstanding conjecture. That being said, it is interesting to note that unlike the case $\alpha=0$
where Conjecture \ref{conj:bd} states that $M_T$ and $\mathbb{E}[X_T]$ are asymptotically the same
for every tree $T$ of order $n$, this is provably not so in many cases of the partial information setting.
Namely, there are cases of trees $T$ and partial labellings $T(S)$  
such that ${\mathbb E}[X_{T(S)}]$ is significantly smaller than $M_{T(S)}$, as the next theorem shows.
Recall that a {\em split} in a tree $T$ is an edge which is not incident to a leaf. Removing a split
from $T$ separates it to two subtrees, as if $e=uv$ is a split, then after removal there are two
tree components $T_u$ and $T_v$, where $u$ is a non-leaf in one component and $v$ is a non-leaf in the other.
Both $u$ and $v$ have degree $2$ in these components so they can be contracted to an edge which we call the {\em join edge} of the split, and hence after contraction, the components
$T_u$ and $T_v$ become proper (phylogenetic) trees.
Note that a split can be identified with the partition $(\L(T_u),\L(T_v))$ of $\L(T)$.

\begin{theorem}\label{t:cat}
	Let $0 < \alpha < 1$. Let $T$ be a tree of order $n$ and let $T(S)$ be a partial labeling where $|S|=\alpha n$. Suppose there exists a split $e=uv$ such that $\L(T_u)=S$ and $T_v$ is
	a caterpillar where the joint edge of $T_v$ is an endpoint edge of the caterpillar. Then
	$$
	M_{T(S)} - {\mathbb E}[X_{T(S)}] = \Theta(n^4)\;.
	$$
\end{theorem}
We note that there are many $T(S)$ which satisfy Theorem \ref{t:cat}. In fact, let $T_1$ be any
labeled tree of order $\alpha n$ and let $T_2=CAT_{n(1-\alpha)}$ where $T_2$ is unlabeled. Split
some edge of $T_1$ introducing a new vertex $u$ and split
some endpoint edge of $T_2$ introducing a new vertex $v$ and connect $u$ and $v$. The resulting $T(S)$
satisfies Theorem \ref{t:cat}. We emphasize that Theorem \ref{t:cat} does not undermine the Badelt-Dress conjecture. In fact, the statement of the theorem is reminiscent of the fact that for the related measure of triplet distance for rooted trees, total incompatibility can be reached.

In Section 3 we develop a form for expressing ${\mathbb E}[X_{T(S)}]$ and use it to
prove our result concerning the minimum expectation, namely Theorem \ref{t:minexp-n}.
In Section 4 we prove our bounds for maximum expectation, namely Theorem \ref{t:maxexp-n}.
The proof uses the form developed in Section 3 but requires considerable more effort in order to exploit it for a good upper bound. In Section 5 we prove Theorem \ref{t:cat}.

\section{${\mathbb E}[X_{T(S)}]$ and $minexp(n,\alpha)$}

Given a partial labeling $T(S)$  with $|S|=k=\alpha n$, we express
${\mathbb E}[X_{T(S)}]$ using certain parameters of the tree involving pairs and triples of
taxa from $[k]$. Although our approach gives an exact expression for ${\mathbb E}[X_{T(S)}]$ it is a
nontrivial task to determine which tree structures and partial assignments thereof optimize that expression.

\subsection{A formula for ${\mathbb E}[X_{T(S)}]$}

Recall that $T$ is a tree on $n$ leaves, $S \subset \L(T)$ with $|S|=k= \alpha n$, the taxa
in $[k]$ are bijectively assigned to $S$, and $T(S)$ denotes this partial labeling.
The set of all $(n-k)!$ completions of $T(S)$ is viewed as a symmetric probability space.
Let $T_0$ be an arbitrary fixed completion of $T(S)$.
Then, $X_{T(S)}=qd(T_0,T^*)$ where $T^*$ is a randomly chosen completion.
We can then write $X_{T(S)}$ as the sum of $\binom{n}{4}$ indicator random variables, one for each
quartet $q \in Q(T_0)$ and we denote this indicator variable by $X_T(q)$,
where $X_T(q)=0$ if $q$ is satisfied by $T^*$.
Indeed observe that
$$
X_{T(S)}=qd(T_0,T^*)=\sum_{q \in \Q(T_0)} X_T(q)\;.
$$
We can split these indicator variables into $5$ types, according to the number of elements of $[k]$ that $q$ contains. We say that an indicator variable $X_T(q)$ is of type $i$ if $q$ contains $i$ elements of $[k]$. We next determine ${\mathbb E}[X_T(q)]$ for the various types.

{\bf Type-$4$ variables}: In this case $X_T(q) \equiv 0$ since all four taxa of $q$ are in the same position in both $T^*$ and $T_0$, so $q$ is satisfied by $T^*$.

{\bf Type-$0$ variables}: In this case ${\mathbb E}[X_T(q)]=\frac{2}{3}$ since each of the three possible quartets on
the elements of $q$ is equally likely to be in $\Q(T^*)$, so $q$ is satisfied by $T^*$ with probability $\frac{1}{3}$.

{\bf Type-$1$ variables}: Suppose $q=[ax|yz]$ where $a \in [k]$.
If we are given the set $P \subset \L(T) \setminus S$ of three leaves of $T^*$
occupying $\{x,y,z\}$ (but not told which of them goes to each element of $P$) each of the three quartets $[ax|yz],[ay|xz],[az|xy]$ is equally likely in $T^*$. Hence, given $P$, $q$ is satisfied by $T^*$ with probability $\frac{1}{3}$.
As this holds for any $P$, we have that ${\mathbb E}[X_T(q)]=\frac{2}{3}$.

{\bf Type-$2$ variables}:
Let $a,b \in [k]$ and consider the sum of all type-$2$ variables involving $a,b$, denoting it by
$X_T(\{a,b\})$.
Let $N_{a,b}$ be the number of pairs of leaves in $\L(T)\setminus S$ such that
if we take two taxa $x,y$ in $[n] \setminus [k]$ and assign $x,y$ to the pair of leaves, then
in the resulting quartet, $a$ and $b$ are not on the same side.
Then $\binom{n-k}{2}-N_{a,b}$ is the number of pairs of leaves in $\L(T)\setminus S$ such that
if we take two taxa $x,y$ in $[n] \setminus [k]$ and assign $x,y$ to the pair of leaves, then
in the resulting quartet, $a$ and $b$ are on the same side.
For convenience, let $p_{a,b} = N_{a,b}/\binom{n-k}{2}$ and so we can view $p_{a,b}$ as the probability that if we take a random pair of leaves in $\L(T)\setminus S$ and assign them to elements of $[n] \setminus [k]$, then in the resulting quartet, the pair $a,b$ will not be on the same side.

Given a type-$2$ variable $X_T(q)$ with $q=[ax|by]$, the probability that $q$ is not satisfied by $T^*$
is $(1-p_{a,b})+p_{a,b}/2=1-p_{a,b}/2$ since with probability $1-p_{a,b}$
in the quartet induced by the taxa of $q$ in $T^*$ the pair $a,b$ will be on the same side
(and then $q$ is surely not satisfied by $T^*$), and with probability $p_{a,b}$ they will
not be on the same side and then $q$ is satisfied by $T^*$ with probability $\frac{1}{2}$,
as the two quartets $[ax|by]$ and $[ay|bx]$ are equally likely.
On the other hand, given a type-$2$ variable $X_T(q)$ with $q=[ab|xy]$, the probability
that $q$ is not satisfied by $T^*$ is clearly $p_{a,b}$.
Summing over all type-$2$ variables involving $a,b$, we have that
$$
{\mathbb E}[X_T(\{a,b\})]=N_{a,b}(1-p_{a,b}/2)+(\binom{n-k}{2}-N_{a,b})p_{a,b}=
\binom{n-k}{2}(2p_{a,b}-1.5p_{a,b}^2)\;.
$$

{\bf Type-$3$ variables}:
Let $a,b,c \in [k]$ and consider the sum of all type-$3$ variables involving $a,b,c$, denoting it by
$X_T(\{a,b,c\})$.
Let $N_{ab|c}$ be the number of leaves in $\L(T)\setminus S$ such that if we assign that leaf a taxa $x$ from $[n] \setminus [k]$ the resulting quartet on $a,b,c,x$ is $[ab|cx]$.
Similarly define $N_{ac|b}$ and $N_{bc|a}$.
Observe that $N_{ab|c}+N_{ac|b}+N_{bc|a}=n-k$, so
by dividing by $n-k$ we can define $p_{ab|c}$, $p_{ac|b}$ and $p_{bc|a}$ accordingly  where $p_{ab|c}+p_{ac|b}+p_{bc|a}=1$.

Given a type-$3$ variable $X_T(q)$ with $q=[ab|cx]$, the probability that $q$ is not satisfied by $T^*$ is $p_{ac|b}+p_{bc|a}$.
Given a type-$3$ variable $X_T(q)$ with $q=[ac|bx]$, the probability that $q$ is not satisfied by $T^*$ is $p_{ab|c}+p_{bc|a}$.
Given a type-$3$ variable $X_T(q)$ with $q=[bc|ax]$, the probability that $q$ is not satisfied by $T^*$ is $p_{ab|c}+p_{ac|b}$.
Summing over all type-$3$ variables involving $a,b,c$, we have that
\begin{align*}
{\mathbb E}[X_T(\{a,b,c\})] & = N_{ab|c}(p_{ac|b}+p_{bc|a})+ N_{ac|b}(p_{ab|c}+p_{bc|a})+N_{bc|a}(p_{ab|c}+p_{ac|b})\\
& = 2(n-k)(p_{ab|c}p_{ac|b}+p_{ab|c}p_{bc|a}+p_{ac|b}p_{bc|a})\;.
\end{align*}

We therefore obtain an exact expression for ${\mathbb E}[X_{T(S)}]$ expressed in terms of the various $p_{a,b}$ for unordered pairs $a,b \in [k]$ and
the various $p_{ab|c}, p_{ac|b}, p_{bc|a}$ for unordered triples $a,b,c \in [k]$ which is:

\begin{align}
\frac{2}{3}\binom{n-k}{4}+ \frac{2}{3}k\binom{n-k}{3}+ \sum_{a,b \in [k]} \binom{n-k}{2}(2p_{a,b}-1.5p_{a,b}^2)+ \nonumber\\
\sum_{a,b,c \in [k]} 2(n-k) (p_{ab|c}p_{ac|b}+p_{ab|c}p_{bc|a}+p_{ac|b}p_{bc|a})\;. \label{e:1}
\end{align}
For $k=\alpha n$ we obtain
\begin{align}
{\mathbb E}[X_{T(S)}] & = \left(\frac{2}{3}(1-\alpha)^4+\frac{8}{3}\alpha(1-\alpha)^3\right)\binom{n}{4} \nonumber\\
& + (1-\alpha)^2\binom{n}{2}\sum_{a,b \in [k]}(2p_{a,b}-1.5p_{a,b}^2) \label{e:2}\\
& + 2(1-\alpha)n\sum_{a,b,c \in [k]}(p_{ab|c}p_{ac|b}+p_{ab|c}p_{bc|a}+p_{ac|b}p_{bc|a})+o(n^4)\;. \nonumber
\end{align}

\subsection{$minexp(n,\alpha)$}

We determine $minexp(n,k/n)$ for every $k=1,\ldots,n-1$ and hence
determine $minexp(n,\alpha)$ asymptotically for every $\alpha \in [0,1]$
thereby proving Theorem \ref{t:minexp-n}.
Our goal is to minimize (\ref{e:1}) over all $T(S)$ with
$|S|=k$. Clearly if $T(S)$ has the property that $p_{a,b}=0$ for all $a,b \in [k]$ and if we could have that two out of the three $p_{ab|c},p_{ac|b},p_{bc|a}$ are zero
for all triples $a,b,c \in [k]$, then we have minimized ${\mathbb E}[X_{T(S)}]$.
Indeed, there are many cases where this can be attained.
Recalling the notion of a split from Section 2, suppose that $T$ is a tree of order $n$
and there is a split $e=uv$ such that in the resulting trees $T_u$ and $T_v$ we have $|\L(T_u)|=k$
and $|\L(T_v)|=n-k$. One example of a tree $T$ for which this holds for all $k=1,\ldots,n-1$ is
$CAT_n$, but for a given $k$, one can construct many other trees with this property.
Now, assume that $S=\L(T_u)$, so indeed $|S|=k$ and consider $T(S)$.
It is immediate to verify that $T(S)$ has the property that $p_{a,b}=0$ for all $a,b \in [k]$.
Indeed, in any quartet of type-$2$ involving $a,b$, we have that $a,b$ will be on the same side in any completion
of $T(S)$. Furthermore,
precisely two out of the three $p_{ab|c},p_{ac|b},p_{bc|a}$ are zero and the third equals $1$ for all triples $a,b,c \in [k]$, as in any quartet of type-$3$ involving the three of them, it holds that in all completions either $a,b$ are always
on the same side (so $p_{ab|c}=1$) or else $a,c$ are on the same side (so $p_{ac|b}=1$) 
or else $b,c$ are on the same side (so $p_{bc|a}=1$).
Hence, we have that $T(S)$ is as small as possible and obtain from (\ref{e:1}) that
$$
minexp(n,k/n) = \frac{2}{3}\binom{n-k}{4}+ \frac{2}{3}k\binom{n-k}{3}\;.
$$
Consequently, using (\ref{e:2}) we obtain that
$$
minexp(n,\alpha) = \left(\frac{2}{3}(1-\alpha)^4+\frac{8}{3}\alpha(1-\alpha)^3\right)\binom{n}{4} + o(n^4)\;.
$$
For $\alpha = \frac{1}{2}$ this gives
$$
minexp(n,\frac{1}{2}) = \frac{5}{24}\binom{n}{4}(1+o(1))\;.
$$
This proves Theorem \ref{t:minexp-n}.
\qed

\section{$maxexp(n,\alpha)$}

We can use (\ref{e:1}) to obtain an upper bound for $maxexp(n,k/n)$ but, as we shall see, unlike the case of $minexp(n,k/n)$, this bound cannot be realized by a construction.
Consider the expressions
$r(a,b)= 2p_{a,b}-1.5p_{a,b}^2$ and $r(a,b,c) = p_{ab|c}p_{ac|b}+p_{ab|c}p_{bc|a}+p_{ac|b}p_{bc|a}$.
Recall that $r(a,b)$ is maximized when $p_{a,b}=\frac{2}{3}$
and $r(a,b,c)$ is maximized when $p_{ab|c}=p_{ac|b}=p_{bc|a}=\frac{1}{3}$.
We therefore have that $r(a,b) \le \frac{2}{3}$ and $r(a,b,c) \le \frac{1}{3}$.
Consequently, we obtain from (\ref{e:1}) that
$$
maxexp(n,k/n) \le \frac{2}{3}\binom{n-k}{4}+\frac{2}{3}k\binom{n-k}{3}+\frac{2}{3}\binom{k}{2}\binom{n-k}{2}
+\frac{2}{3}\binom{k}{3}(n-k)\;.
$$
Using (\ref{e:2}) we obtain that
$$
maxexp(n,\alpha) \le \left(\frac{2}{3}-\frac{2\alpha^4}{3}\right)\binom{n}{4}     + o(n^4)
$$
which for $\alpha = \frac{1}{2}$ this gives
$$
maxexp(n,\frac{1}{2}) \le \frac{5}{8}\binom{n}{4}(1+o(1))\;.
$$
Unfortunately, it is impossible to {\em simultaneously} achieve $r(a,b)=\frac{2}{3}$ for all pairs $a,b$ and $r(a,b,c)=\frac{1}{3}$ for all triples $a,b,c$. There is, however, a construction which does achieve
these optimal bounds for many pairs and many triples, so we use this construction as our lower bound
for $maxexp(n,\alpha)$. 

\subsection{Lower bound for $maxexp(n,\alpha)$}
Take three copies of the caterpillar $CAT_{n/3+1}$, denoting them by $C_1,C_2,C_3$.
Denote the leaves of $C_i$ ordered from left to right by $c_{i,0},\ldots,c_{i,n/3}$.
Assign the taxa $\{1,..,\alpha n/3\}$ to the leaves $c_{1,1},\ldots,c_{1,\alpha n/3}$.
Assign the taxa $\{\alpha n/3+1,...,2\alpha n/3\}$ to the leaves $c_{2,1},\ldots,c_{2,\alpha n/3}$.
Assign the taxa $\{2\alpha n/3+1,\ldots,\alpha n\}$ to the leaves $c_{3,1},\ldots,c_{3,\alpha n/3}$.
Now create a tree $H$ by identifying the $c_{i,0}$ for $i=1,2,3$, so $H$ is a ``$3$-way caterpillar''.
As $\alpha n$ of the leaves of $H$ are assigned with the taxa set $[k]=[\alpha n]$, we denote this partial labeling by $H(S)$.
We compute (\ref{e:1}) for $H(S)$. The number of pairs $a,b$ of elements of $[k]$
where $a$ and $b$ are in distinct $C_i$ is $3(\alpha n/3)^2=\alpha^2 n^2/3$ and for such pairs we indeed have $p_{a,b}=(2/3)+o(1)$ which, recall, makes $r(a,b)$ asymptotically reach its maximum value $2/3$.
There are $3\binom{\alpha n/3}{2} = (1+o(1))\alpha^2n^2/6$ pairs $a,b$ of elements of $S$
where $a$ and $b$ are in the same $C_i$. For such pairs we have $p_{a,b}=(4/9)+o(1)$
as for $a$ and $b$ to be in different sides of a quartet we must choose an unlabeled leaf from $C_i$
and another unlabeled leaf not from $C_i$. Hence, $r(a,b)=(16/27)+o(1)$ in this case.
There are $\alpha^3 n^3/27$ triples $a,b,c \in S$ where each is in a distinct $C_i$ and in this case we have
$p_{ab|c}=p_{ac|b}=p_{bc|a}=(1/3)+o_n(1)$ so $r(a,b,c)=\frac{1}{3}+o(1)$ and asymptotically reaches its maximum value.
There are $3\binom{\alpha n/3}{3}=(1+o(1))\alpha^3n^3/54$ triples $a,b,c \in S$ where all three are in the same $C_i$ and in this case we have that $\{p_{ab|c},p_{ac|b},p_{bc|a}\}=\{0,(1/3)+o(1),(2/3)+o(1)\}$ (which variable gets which value depends on the order of $a,b,c$ inside the $C_i$). Then, $r(a,b,c)=(2/9)+o(1)$ in this case.
Finally, there are $2\alpha n\binom{\alpha n/3}{2}=(1+o(1))\alpha^3n^3/9$ triples $a,b,c \in S$ where one of them is in some $C_i$ and the other two are in another $C_i$. Here it also holds that  $\{p_{ab|c},p_{ac|b},p_{bc|a}\}=\{0,(1/3)+o(1),(2/3)+o(1)\}$ and again $r(a,b,c)=(2/9)+o(1)$.
Plugging these values in (\ref{e:2}) we obtain
\begin{align*}
{\mathbb E}[X_{H(S)}] & = \left(\frac{2}{3}(1-\alpha)^4+\frac{8}{3}\alpha(1-\alpha)^3\right)\binom{n}{4}(1+o(1))\\
& +
(1-\alpha)^2\binom{n}{2}\left( \frac{2}{3}\cdot \frac{\alpha^2 n^2}{3}+\frac{16}{27}\cdot\frac{\alpha^2 n^2}{6}\right)(1+o(1))\\
& + 2(1-\alpha)n\left(\frac{1}{3}\cdot\frac{\alpha^3n^3}{27}+\frac{2}{9}\cdot\frac{\alpha^3n^3}{54}+
\frac{2}{9}\cdot\frac{\alpha^3n^3}{9}\right)(1+o(1))\\
& = \binom{n}{4}\left(\frac{2}{3}-\frac{10\alpha^4+32\alpha^3+12\alpha^2}{81}\right)(1+o(1))
\end{align*}
We therefore have
$$
maxexp(n,\alpha) \ge \binom{n}{4}\left(\frac{2}{3}-\frac{10\alpha^4+32\alpha^3+12\alpha^2}{81}\right)(1+o(1))
$$
and in particular,
$$
maxexp(n,\frac{1}{2}) \ge \frac{371}{648}\binom{n}{4}(1+o(1)) \ge 0.572\binom{n}{4}(1+o(1))\;.
$$

\subsection{Upper bound for $maxexp(n,\alpha)$}
Consider the expressions 
\begin{align*}
f(T(S)) & = \sum_{a,b,c \in [k]} (p_{ab|c}p_{ac|b}+p_{ab|c}p_{bc|a}+p_{ac|b}p_{bc|a})=\sum_{a,b,c \in [k]}r(a,b,c)\\
g(T(S)) & = \sum_{a,b \in [k]} (2p_{a,b}-1.5p_{a,b}^2) = \sum_{a,b \in [k]} r(a,b)\;.
\end{align*}
We first present our building block for upper-bounding these quantities.

Recall from Section 2 that for $A \subseteq \L(T)$,
the topological subtree $T|_A$ is the tree obtained by removing the leaves $\L(T)\setminus A$,
the paths leading exclusively to them,  and contracting vertices with degree two.
Now suppose we are given a partial labeling $T(S)$ where the leaves in $S$ are labeled with the taxa set $[k]$. Suppose $2 \le r \le k$ is fixed and let $A \subseteq S$ be fixed with $|A|=r$.
Consider $T|_A$ which is a labeled tree of order $r$ where its $r$ leaves are labeled by the taxa from $[k]$ assigned to $A$. Also observe that $T|_A$ has $2r-3$ edges and denote the set of edges of $T|_A$
by $F$. We can partition the leaves in $\L(T) \setminus S$ (namely, the unlabeled leaves of $T(S)$)
into $|F|=2r-3$ parts as follows. Recall that each $e \in F$ corresponds to some contracted path $e^*$
of $T$. Now let $\ell \in \L(T) \setminus S$ and consider the path in $T$ connecting $\ell$ to some
arbitrary leaf of $A$. So, this path starts at $\ell$ and then hits for the first time some path $e^*$.
Then we place $\ell$ in the set $X_e$. Notice that for every $\ell \in \L(T) \setminus S$, the set
$X_e$ in which it is placed is well-defined, and $\{X_e\,:\, e \in F\}$ is a partition of
$\L(T) \setminus S$ (some parts may be empty).

Consider some triple $a,b,c \in A$ and the variable $p_{ab|c}$.
We claim that we can determine $p_{ab|c}$ from the sizes of the $X_e$'s. Indeed, suppose $x$ is some
taxa assigned to some leaf $\ell \in \L(T) \setminus S$. Then $[ab|cx]$ is a quartet of $T$ if and only if
the following holds: $\ell \in X_e$ where the path from $c$ to $e$ in $T|_A$ does not use any edge
of the path from $a$ to $b$ in $T|_A$. Thus, if $F(ab|c) \subset F$ is the union of all edges of $F$
such that the path leading from $c$ to each of them does not use  any edge
of the path from $a$ to $b$ in $T|_A$, then, denoting $x_e=|X_e|/(n-k)$ we have
$$
p_{ab|c} = \frac{\sum_{e \in F(ab|c)} |X_e|}{|\L(T) \setminus S|}= \sum_{e \in F(ab|c)} x_e\;.
$$

\begin{figure}
	\includegraphics[scale=0.6,trim=-190 340 0 40, clip]{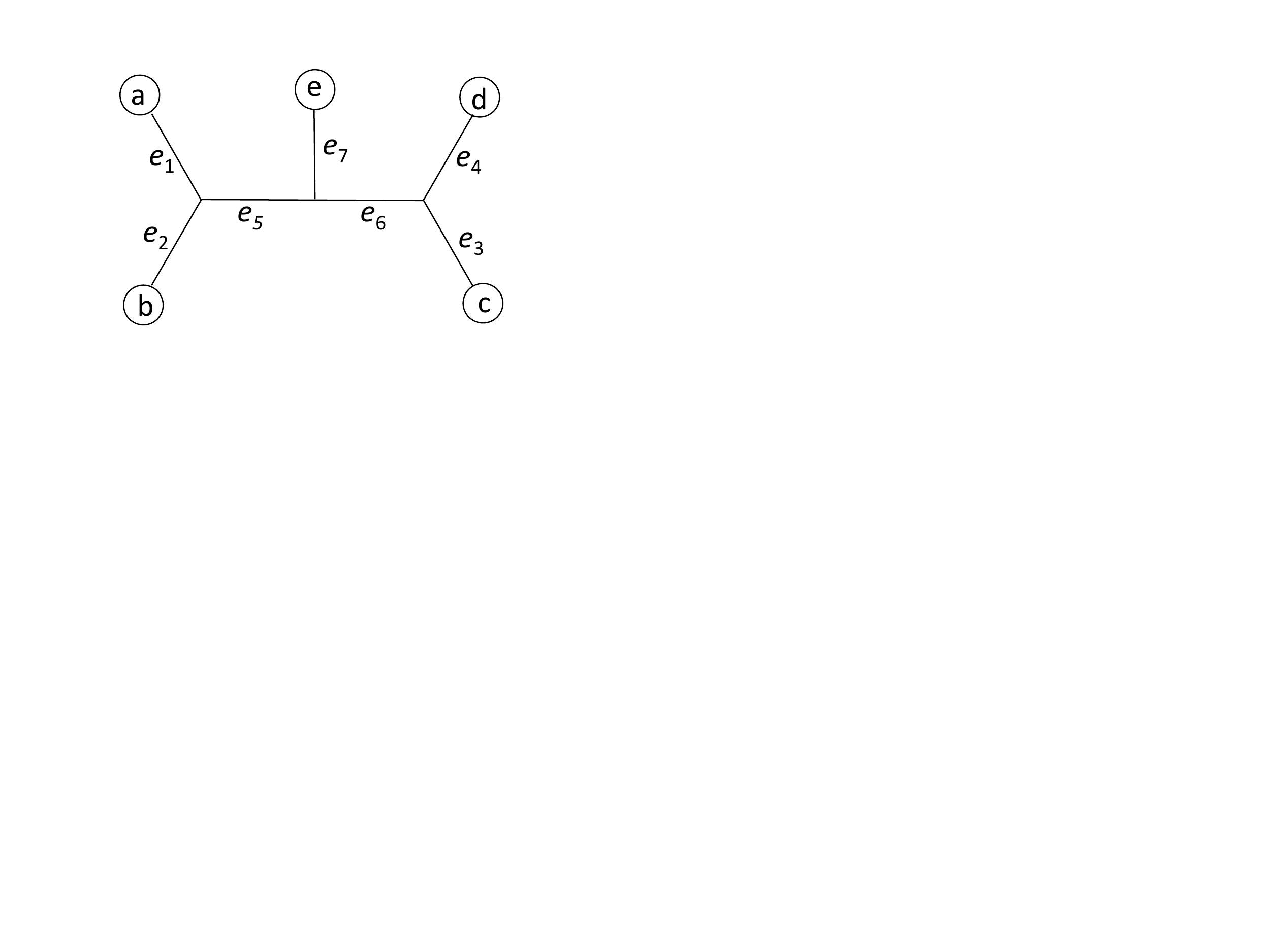}
	\caption{The quintet with $A=\{a,b,c,d,e\}$ and $F=\{e_1,\ldots,e_7\}$.}
	\label{f:quintet}
\end{figure}

For our bound for $f(T(S))$ we will use $r=5$. Indeed this is the largest $r$ such that $T|_A$ is
unique, as there is only one tree of order $5$ which is the quintet, or, equivalently, $CAT_5$.
(For larger $r$ we must consider all possible tree structures of order $r$ and for each of them derive and solve an optimization problem; the gain is marginal compared with $r=5$ and the size of the
problems gets very large.) So, assume $A=\{a,b,c,d,e\}$ and that $T|_A$ is given in Figure \ref{f:quintet} where we have labeled the edges of $F=\{e_1,\ldots,e_7\}$.

In Table \ref{table:var-1} we have listed the
various variables and the corresponding expressions we obtain in this case. For shorthand, we defined $x_i=x_{e_i}$ so the variables are $x_1,\ldots,x_7$.
Bearing in mind that $x_1+\cdots+x_7=1$ we can now optimize the sum of the various expressions
$r(u,v,w) = p_{uv|w}p_{uw|v}+p_{uv|w}p_{vw|u}+p_{uw|v}p_{vw|u}$ for $u,v,w \in A=\{a,b,c,d,e\}$.
In other words, we maximize
$$
r(A)=\sum_{u,v,w \in A} r(u,v,w) \qquad\qquad {\rm s.t.} ~~~ x_1+x_2+x_3+x_4+x_5+x_6+x_7=1\;.
$$
From Table \ref{table:var-1} this amounts to finding the global maximum of a quadratic polynomial in $6$ variables
$x_1,\ldots,x_6$ and its maximum in attained when $x_1=x_2=x_3=x_4=\frac{7}{33}$, $x_5=x_6=0$ and
$x_7=\frac{5}{33}$. In that point the  maximum is $\frac{98}{33}$.

\begin{table}
	\vspace{1cm}
	\centering
	\begin{tabular}{|c|c|}
		\hline
		variable & expression in terms of the $x_i$'s\\
		\hline
		$p_{ab|c}$, $p_{ab|d}$, $p_{ab|e}$  & $x_3+x_4+x_5+x_6+x_7$\\
		\hline
		$p_{cd|a}$, $p_{cd|b}$, $p_{cd|e}$  & $x_1+x_2+x_5+x_6+x_7$\\
		\hline
		$p_{ae|c}$, $p_{ae|d}$, $p_{be|c}$, $p_{be|d}$  & $x_3+x_4+x_6$\\
		\hline
		$p_{ce|a}$, $p_{ce|b}$, $p_{de|a}$, $p_{de|b}$  & $x_1+x_2+x_5$\\
		\hline
		$p_{bc|a}$, $p_{bd|a}$, $p_{be|a}$ & $x_1$\\
		\hline
		$p_{ac|b}$, $p_{ad|b}$, $p_{ae|b}$ & $x_2$\\
		\hline
		$p_{ad|c}$, $p_{bd|c}$, $p_{de|c}$ & $x_3$\\
		\hline
		$p_{ac|d}$, $p_{bc|d}$, $p_{ce|d}$ & $x_4$\\
		\hline
		$p_{ac|e}$, $p_{ad|e}$, $p_{bc|e}$, $p_{bd|e}$ & $x_7$\\
		\hline
	\end{tabular}
	\caption{The variables whose values can be derived in terms of the $x_i$'s from the quintet of Figure \ref{f:quintet}. Variables having the same expression are grouped together.}
	\label{table:var-1}
\end{table}

Returning now to $f(T(S))$ we observe that the number of $A \subset S$ with $|A|=5$ is
$\binom{k}{5}$ and each triple $\{a,b,c\} \subset S$ occurs in $\binom{k-3}{2}$ such $A$.
As we have proved that $r(A) \le \frac{98}{33}$ for each $A$, this implies that
$$
f(T(S)) \le \frac{98}{33}\frac{\binom{k}{5}}{\binom{k-3}{2}}=\frac{49}{990}k(k-1)(k-2)\;.
$$

We can use $T|_A$ of Figure \ref{f:quintet} to upper bound $g(T(S))$ as well, but in this case, the
values of the variables do not solely depend on the $x_i$'s. Indeed, consider the variable $p_{a,b}$
and recall that it corresponds to the probability that a randomly chosen pair of leaves in
$\L(T) \setminus S$ forms together with $a,b$ a quartet in which $a$ and $b$ are on different sides.
So, $1-p_{a,b}$ corresponds to $a,b$ remaining on the same side, or stated otherwise, the probability that if we chose two leaves form $\L(T) \setminus S$, the path connecting them will not pass through the path connecting $a$ and $b$. Thus, considering Figure \ref{f:quintet}, we have that
$1-p_{a,b} = (x_3+x_4+x_5+x_6+x_7)^2 + d_1 +d_2$ where $d_i$ is the probability that a
chosen pair of leaves from $X_i$ will be connected with a path that does not intersect $e_i^*$.
But observe  that unlike the case for $f(T(S))$ and Table \ref{table:var-1} where the expressions only
involve the $x_i$'s, here the expressions involve the $d_i$'s as well, and the only thing we know
about them is that $0 \le d_i \le x_i^2$. In Table \ref{table:var-2} we have listed the
various variables and the corresponding expressions we obtain in this case.

\begin{table}
	\vspace{1cm}
	\centering
	\begin{tabular}{|c|c|}
		\hline
		variable & expression in terms of the $x_i$'s and $d_i$'s\\
		\hline
		$p_{a,b}$ & $1-(x_3+x_4+x_5+x_6+x_7)^2 - d_1 -d_2$\\
		\hline
		$p_{a,c}$ & $1-x_4^2-x_2^2-x_7^2-d_5-d_6-d_1-d_3$\\
		\hline
		$p_{a,d}$ & $1-x_3^2-x_2^2-x_7^2-d_5-d_6-d_1-d_4$\\
		\hline
		$p_{a,e}$ & $1-(x_3+x_4+x_6)^2-x_2^2-d_1-d_5-d_7$\\
		\hline
		$p_{b,c}$ & $1-x_1^2-x_4^2-x_7^2-d_5-d_6-d_2-d_3$\\
		\hline
		$p_{b,d}$ & $1-x_1^2-x_3^2-x_7^2-d_5-d_6-d_2-d_4$\\
		\hline
		$p_{b,e}$ & $1-(x_3+x_4+x_6)^2-x_1^2-d_2-d_5-d_7$\\
		\hline
		$p_{c,d}$ & $1-(x_1+x_2+x_5+x_6+x_7)^2-d_3-d_4$\\
		\hline
		$p_{c,e}$ & $1-(x_1+x_2+x_5)^2-x_4^2-d_3-d_6-d_7$\\
		\hline
		$p_{d,e}$ & $1-(x_1+x_2+x_5)^2-x_3^2-d_4-d_6-d_7$\\
		\hline
	\end{tabular}
	\caption{The variables of the form $p_{u,v}$ whose values, up to $o(1)$, can be derived in terms of the $x_i$'s and $d_i$'s from the quintet of Figure \ref{f:quintet}.}
	\label{table:var-2}
\end{table}
We can now optimize the sum of the various expressions
$r(u,v) = 2p_{u,v}-1.5p_{u,v}^2$ for $u,v \in A=\{a,b,c,d,e\}$.
In other words, we wish to maximize
$$
\sum_{u,v \in \{a,b,c,d,e\}} r(u,v) \qquad\qquad {\rm s.t.} ~~~ x_1+x_2+x_3+x_4+x_5+x_6+x_7=1,
~~~ 0 \le d_i \le x_i^2 ~{\rm for }~ i=1,\ldots,7\;.
$$
There is no simple analytic solution to this nonlinear optimization problem (notice that it involves the sum of polynomials of degree at most $4$ and that it involves inequalities, as opposed to the case of $f(T(S))$ which required optimizing a single quadratic polynomial)
and we can only solve it numerically using a nonlinear optimization algorithm, such as NLPSolve of Maple
\footnote{Maple code of our program available at \url{https://www.dropbox.com/s/va6dwf3yw773vyx/quartets-partial.zip?dl=0}}. Using that method, we obtain that the maximum is (slightly less than) $6.6316$
with $x_1 = x_4 = 0.08134$, $x_2=x_3=0.39336$, $x_7=0.05059$, $x_5=x_6=0$, $d_1=d_4=0$,
$d_2=d_3 = 0.12914$, $d_5=d_6=d_7=0$. Compare this maximum to the one we can obtain from the trivial
upper bound $r(u,v) \le \frac{2}{3}$ which would give a bound of $6.666...$ (as there are $10$ terms in the sum to maximize). So, while for the case of $f(T(S))$ we obtained a significant improvement
(expressed analytically as a rational number), here the improvement is rather marginal (and we cannot express the bound analytically).
In any case, we have
$$
g(T(S)) \le 6.6316 \frac{\binom{k}{5}}{\binom{k-2}{3}}=0.33158 k(k-1)\;.
$$
Plugging our bounds for $f(T(S))$ and $g(T(S))$ in (\ref{e:2}) we obtain that
$$
maxexp(n,\alpha) \le \left(\frac{2}{3}(1-\alpha)^4+\frac{8}{3}\alpha(1-\alpha)^3+3.98\alpha^2(1-\alpha)^2+
\frac{392}{165}\alpha^3(1-\alpha)\right)\binom{n}{4}     + o(n^4)
$$
$$
= \left(\frac{2}{3}-\frac{653\alpha^4+414\alpha^3+33\alpha^2}{1650}\right)\binom{n}{4}+o(n^4)
$$
which for $\alpha = \frac{1}{2}$ gives
$$
maxexp(n,\textstyle \frac{1}{2}) \le 0.606\binom{n}{4}(1+o(1))\;.
$$
If one only uses the easier bound $g(T(S)) \le k(k-1)/3$ then the constant $3.98$ in the bound for $maxexp(n,\alpha)$ can be replaced with $4$ and the constant $0.606$ in the bound for
$maxexp(n,\frac{1}{2})$ can be replaced with $\frac{267}{440} < 0.607$.
We have thus completed the proof of Theorem \ref{t:maxexp-n}. \qed

\section{${\mathbb E}[X_{T(S)}]$ versus $M_{T(S)}$}

In  this section we prove Theorem \ref{t:cat}.
So we assume that $T$ is a tree of order $n$ with a split $e=uv$ such that $\L(T_u)=S$ and $T_v$ is
a caterpillar (where the join edge of $T_v$ is an endpoint edge of the caterpillar) and our partial labeling is $T(S)$. Notice that $T(S)$ is a special case of the partial labellings that minimize ${\mathbb E}[X_{T(S)}]$ as proved in Subsection 3.2 (the special case is that
$T_v$ is a caterpillar and for minimizing ${\mathbb E}[X_{T(S)}]$ this is not required).
In particular, by Theorem \ref{t:minexp-n},
\begin{equation}\label{e:xts}
{\mathbb E}[X_{T(S)}]=
\left(\frac{2}{3}(1-\alpha)^4+\frac{8}{3}\alpha(1-\alpha)^3\right)\binom{n}{4}(1+o(1))\;.
\end{equation}

We proceed to lower-bound $M_{T(S)}$. Recall that the vertices of $S$ are labeled with the taxa set $[k]$
where $k=\alpha n$.
We construct two completions of $T(S)$, denoted by $T_1$ and $T_2$ as follows. Suppose that
$\L(T_v)=\{\ell_1,\ldots,\ell_{n-k}\}$ where leaf $\ell_j$ is closer to the split than leaf $\ell_{j+1}$.
A completion assigns the taxa set $\{k+1,\ldots,n\}$ to
$\L(T_v)$. In $T_1$ we assign taxa $k+j$ to leaf $\ell_j$ for $j=1,\ldots,n-k$.
We next describe $T_2$. For ease of exposition we separate the cases $\alpha > 1/3$ and $\alpha \le 1/3$.

Consider first the case $\alpha > 1/3$.
In $T_2$ we reverse the ordering with respect to $T_1$ and assign taxa $k+j$ to leaf $\ell_{n-k+1-j}$.
We compute $qd(T_1,T_2)$. Quartets of type $i=0,2,3,4$ of $T_1$ are also quartets in $T_2$
where recall that a quartet is of type $i$ if it contains $i$ taxa from $S$.
On the other hand, each quartet of type $1$ in $T_1$ is not a quartet of $T_2$. Since assume that
the quartet is of the form $[ax|yz]$ in $T_1$. Then $x$ is assigned to a leaf closer to the split
than the leaves assigned to $y$ and $z$. But in $T_2$, the leaf assigned to $z$ will be closer to
the split than the leaves assigned to $y$ and $x$. So, in $T_2$ the quartet would be $[az|yx]$.
It follows that $qd(T_1,T_2)$ equals the number of quartets of type $1$ hence
\begin{equation}\label{e:mts}
M_{T(S)} \ge qd(T_1,T_2)=\alpha n \binom{(1-\alpha)n}{3} = 4\alpha(1-\alpha)^3\binom{n}{4}(1+o(1))\;.
\end{equation}
Simple calculus shows that (\ref{e:mts}) minus (\ref{e:xts}) is $\Theta(n^4)$ for $\alpha > 1/3$.

When $\alpha \le 1/3$ the aforementioned ``reversal labeling'' for $T_2$ produces inferior results than that of a random labeling. We now show how to obtain a more delicate labeling of $T_2$ which is a mixture of a random and a deterministic labeling. Let $1-\alpha \ge \beta \ge 0$ be a parameter. We partition $\L(T_v)$ into three parts. The $\beta n/2$ leaves closest to the split are denoted $B_1$, so $B_1=\{\ell_1,\ldots,\ell_{\beta n/2}\}$.  The last $\beta n/2$ leaves furthest from
the split are denoted $B_2=\{\ell_{(1-\alpha-\beta/2)n+1},\ldots,\ell_{(1-\alpha)n}\}$.
The remaining $(1-\alpha-\beta)n$ leaves in $\L(T_v)$ (the middle leaves of the caterpillar) are denoted by $C$. Let $B=B_1 \cup B_2$ and observe that $|B|=\beta n$, $|C|=(1-\alpha-\beta)n$ and
$B \cup C= \L(T_v)$.
The taxa assigned to $C$ in $T_1$ are randomly permuted in $T_2$, but stay inside $C$ so that any label assigned in $T_1$ to a leaf of $C$ is also assigned in $T_2$ to some leaf of $C$.
The taxa assigned to $B_1$ in $T_1$ are now assigned to $B_2$ in $T_2$ and are also reversed.
Similarly, the taxa assigned to $B_2$ in $T_1$ are now assigned to $B_1$ in $T_2$ and are also reversed.
In particular, for $j=1,\ldots,\beta n/2$, taxa $k+j$ that was assigned in $T_1$ to $\ell_j \in B_1$
is now assigned in $T_2$ to $\ell_{(1-\alpha)n+1-j} \in B_2$ and taxa $n+1-j$ that was assigned in
$T_1$ to $\ell_{(1-\alpha)n+1-j} \in B_2$ is now assigned in $T_2$ to $\ell_j \in B_1$.

Observe that now $qd(T_1,T_2)$ is a random variable as the taxa assigned to $C$ in $T_1$ are randomly permuted in $T_2$. We compute ${\mathbb E}[qd(T_1,T_2)]$ which (obviously) serves as a lower bound
for $M_{T(S)}$. Once again, quartets of type $i=4,3,2$ of $T_1$ are also quartets in $T_2$ so it
remains to consider quartets of type $0$ and type $1$. We will designate such quartets by words
of the form $\{S,B,C\}^4$ where a word of the form $W_1W_2W_3W_4$ represents a quartet of $T_1$
of the form $[x_1x_2|x_3x_4]$ where $x_i \in W_i$. So, for example, $SBCB$ represents quartets
$[x_1x_2|x_3x_4]$ where $x_1 \in S$, $x_2 \in B$, $x_3 \in C$, $x_4 \in B$ (notice that this is a quartet of type-$1$). Recall that we can view $qd(T_1,T_2)$ as the sum of indicator random variables,
each representing a quartet of type $0$ or $1$ in $T_1$ and the probability it is not compatible with $T_2$.
In Table \ref{table:var-3} we list each possible quartet subtype, the number of indicator random variables (i.e. quartets) of this subtype normalized by dividing with $(1+o(1))\binom{n}{4}$, and the probability it is not compatible with $T_2$. Using Table \ref{table:var-3} we immediately obtain
	
\begin{table}
\vspace{1cm}
\centering
\begin{tabular}{|c|c|c|}
			\hline
			subtype & normalized quantity & probability\\
			\hline
			$CCCC$ & $(1-\alpha-\beta)^4$ & $\frac{2}{3}$\\
			\hline
			$BCCC \cup CCCB$ & $4\beta(1-\alpha-\beta)^3$  & $\frac{2}{3}$\\
			\hline
			$BBCC \cup CCBB$ & $3\beta^2(1-\alpha-\beta)^2$ & $0$\\
			\hline
			$BCCB$ & $3\beta^2(1-\alpha-\beta)^2$  & $\frac{1}{2}$\\
			\hline
			$BBCB \cup BCBB \cup BBBC \cup CBBB$ & $4\beta^3(1-\alpha-\beta)$ & $0$\\
			\hline
			$BBBB$ & $\beta^4$ & $0$\\
			\hline
			$SCCC$ & $4\alpha(1-\alpha-\beta)^3$ & $\frac{2}{3}$\\
			\hline
			$SBCC \cup SCCB$ & $12\alpha\beta(1-\alpha-\beta)^2$ & $1$\\
			\hline
			$SBBC \cup SBCB \cup SCBB$ & $12\alpha\beta^2(1-\alpha-\beta)$ & $1$\\
			\hline
			$SBBB$ & $4\alpha\beta^3$ & $1$\\
			\hline
		\end{tabular}
		\caption{The various subtypes of quartets of type $0$ or $1$, the normalized quantity of quartets of each subtype or union of subtypes, and the probability it is not compatible with $T_2$.}
		\label{table:var-3}
\end{table}

\begin{align*}
\frac{1}{\binom{n}{4}(1+o(1))}{\mathbb E}[qd(T_1,T_2)] & = \frac{2}{3}(1-\alpha-\beta)^4\\
& + \frac{8}{3}\beta(1-\alpha-\beta)^3\\
& + \frac{3}{2}\beta^2(1-\alpha-\beta)^2\\
& + \frac{8}{3}\alpha(1-\alpha-\beta)^3\\
& + 12\alpha\beta(1-\alpha-\beta)^2\\
& + 12\alpha\beta^2(1-\alpha-\beta)\\
& + 4\alpha\beta^3\;.
\end{align*}

Simplifying, the right hand side of the last equality is
$$
g(\beta)=\alpha\beta^{2}-\alpha\beta^3-\frac{5}{2}\beta^2+\frac{7}{3}\beta^3+4\alpha\beta+\frac{2}{3}
-4\alpha^2-8\alpha^2\beta+4\alpha^3\beta+\frac{3}{2}\alpha^2\beta^2+\frac{16}{3}\alpha^3-2\alpha^4-
\frac{1}{2}\beta^4
$$
where we have named the expression $g(\beta)$ to emphasize that $\alpha$ is fixed.
Hence using (\ref{e:xts}) our hope is to show that there exists $0 \le \beta \le 1-\alpha$
such that $g(\beta) > \frac{2}{3}(1-\alpha)^4+\frac{8}{3}\alpha(1-\alpha)^3$
as this will prove that
$$
M_{T(S)} - {\mathbb E}[X_{T(S)}] \ge {\mathbb E}[qd(T_1,T_2)]-{\mathbb E}[X_{T(S)}] =\Theta(n^4)
$$
as required. To see that the claim on $g(\beta)$ holds, first observe that
$g(0)=  \frac{2}{3}(1-\alpha)^4+\frac{8}{3}\alpha(1-\alpha)^3$.
As $g$ is differentiable, it suffices to prove that $g'(0) > 0$.
Indeed 
$$
g'(\beta)= 2\alpha\beta-3\alpha\beta^2-5\beta+7\beta^2+4\alpha-8\alpha^2 + 4\alpha^3+3\alpha^2\beta-2\beta^3
$$
so
$$
g'(0)=4\alpha-8\alpha^2 + 4\alpha^3=4\alpha(1-\alpha)^2 > 0\;.
$$
\qed

\end{document}